\documentclass[10pt]{iopart}

\usepackage{iopams}  
\usepackage{graphicx}
\usepackage{color}
\usepackage{cite}
\usepackage{bm}
\usepackage{mathrsfs}
\usepackage{braket}
\usepackage{bbold}
\expandafter\let\csname equation*\endcsname\relax
\expandafter\let\csname endequation*\endcsname\relax
\usepackage{amsmath}
\usepackage[titletoc,title]{appendix}\usepackage{fancyhdr}
\newcommand*\Bell{\ensuremath{\boldsymbol\ell}}

\begin{document}

\title{Modelling toolkit for simulation of maglev devices}

\author{J. Pe\~na--Roche and A. Bad\'{\i}a--Maj\'os$^*$}
\address{Departamento de F\'{\i}sica de la Materia
Condensada \\ and Instituto de Ciencia de Materiales de Arag\'{o}n (ICMA), \\ Universidad de Zaragoza--CSIC, 
C/ Mar\'{\i}a de Luna 3, E-50018 Zaragoza, Spain}
\ead{anabadia@unizar.es; $^*${\rm corresponding author}}
\vspace{10pt}
\begin{indented}
\item[]October 2016
\end{indented}

\begin{abstract}
A stand-alone App\footnote[1]{the {\sc Android} compatible App as well as an Open Virtualization Archive are freely available upon request to the authors}
has been developed, focused on obtaining information about relevant engineering properties of magnetic levitation systems. Our modelling toolkit provides real time simulations of 2D magneto-mechanical quantities for Superconductor/Permanent Magnet structures.
The source code is open and may be customized for a variety of configurations. Ultimately, it relies on the variational statement of the critical state model for the superconducting component and has been verified against experimental data for YBaCuO/NdFeB  assemblies.

On a quantitative basis, the values of the arising forces, induced superconducting currents, as well as a plot of the magnetic field lines are displayed upon selection of an arbitrary trajectory of the magnet in the vicinity of the superconductor. The stability issues related to the cooling process, as well as the maximum attainable forces for a given material and geometry are immediately observed.

Due to
the complexity of the problem, a strategy based on cluster computing, database compression, and real-time post-processing on the device has been implemented.
\end{abstract}

\pacs{03.50.De, 13.40.-f, 74.20.De, 02.30.Yy, 02.70.-c}

\vspace{2pc}
\noindent{\it Keywords}: Magnetic levitation, Critical state model, Superconducting modelling
\vspace{2pc}

\submitto{\SUST}
\ioptwocol
\section{Introduction}
\label{sec:1}
Maglev devices based on permanent magnets (PMs) and superconductors (SCs) constitute one of the most promising large scale applications of High-Temperature Superconductors \cite{ref1_1, ref1_2,ref1_3,ref1_4,ref1_5,ref1_6,ref1_7}.

It is well known that the accurate description of the physical properties of real engineering systems involves non-trivial modelling. \textcolor{black}{Qualitative} and even quantitative predictions of the involved phenomena are relatively simple and may be formulated by elementary Electromagnetics, as far as idealized geometry and simple movements of the magnet (or superconductor) are involved. However, realistic finite-size effects as well as non-trivial displacements are hard to introduce if quantification is required. \textcolor{black}{For instance, if one wants to argue about the influence of the relative position of the PM/SC components previous to the cooling down process of the superconductor, non-trivial path-dependent aspects must be analyzed. As far as available degrees of freedom are at hand, one may wish to analyze how to perform the process so as to achieve the highest possible levitation, suspension or guidance forces as well as reasonable  stiffness values}. Hitherto, a number of more or less elaborate numerical models have been reported that solve 2D or even 3D geometry with a good degree of approximation to the experimental facts \cite{ref2_1,ref2_2,ref2_3,ref3}. Mainly, such models consider hard magnets with constant magnetization and either introduce a power-law $E(J)$ characteristic for the superconductor, or its limiting form, i.e.: the {\em critical state} ansatz \cite{bean}.

Generally speaking, the above mentioned methods have been implemented in the form of sophisticated numerical codes that run on a workstation, demanding certain computing resources and involving moderate times of calculation. In this work, we have confronted the objective of creating a numerical simulation tool that may realize quantitative predictions for Maglev devices in ``real time'', based on the simplest possible computation resource. To be specific, we have targetted the implementation of the already well established physical models in the form of a utility for the more and more popular portable devices. Expectedly, this still unconventional modelling tool will allow to perform quick, but rigorous calculations in a simple manner. In brief, finger tapping and sliding gestures across a graphical screen are enough to define the physical problem and prompt a visual solution that may be straightforwardly used as preliminary design information. 

\textcolor{black}{Mainly, the purpose of this work was to provide a comprehensible data processing tool that may be of help for analyzing the counter-intuitive, but highly relevant, hysteresis effects that occur when the magnet wags around the superconductor. It may be a resource for guiding discussions and physical interpretations in Maglev design (as a complement to the dedicated computer simulation tools).} \textcolor{black}{Also, to a large extent, it has a focus on enlightenment, training and dissemination actions.}

Concerning the portable devices, in spite of the more and more capable systems, mathematical libraries and general purpose software are still scarce, and straight stand-alone simulations are not affordable yet. For this reason, a hybrid strategy that combines pre-processing in a high performance computer and eventual post-processing on the device has been designed.

The article is organized as follows. First (Sec.\ref{sec:2}), we summarize the mathematical statement of the levitation problem. The physical model equations and their formulation via the {\em finite element} approximation will be discussed. Validation against experimental data on typical materials will be presented. Second, in Sec.\ref{sec:3}, we will introduce some concepts on the \textcolor{black}{actual} implementation. This is intended to provide a basic idea about our solution for performing physical simulations in a portable device. \textcolor{black}{Some examples are analysed, as an illustration of the visual output of the App. This gives a sort of guide for the potential end-users of the software.}
Possible extensions of the numerical toolkit will be eventually discussed.

\section{The MagLev problem: physical model}
\label{sec:2}
Below, we present the physical modelling that is used to characterize the system formed by the magnet and superconductor. The detailed background theory may be found in our previous Refs.\cite{ref3,gcs}. Here, we just introduce the essentials for self-consistency.

In brief, the main idea is to solve the electromagnetic problem defined by the positioning of both components. By using conventional terms this may be described:

\begin{enumerate}
\item[(i)] The Permanent Magnet is a magnetic field source that creates a given vector potential around, say ${\bf A}_{0}(x,y,z)$ 
\item[(ii)] The Superconductor responds to the sources through induced electrical current density within its volume ${{\bf J}_{\rm sc}}({x,y,z})$.  
\item[(iii)] ${{\bf J}_{\rm sc}}({x,y,z})$ may be obtained from the Maxwell equations and some specific material law.
\item[(iv)] All the electromechanical properties may be calculated based on ${\bf A}_{0}$ and ${{\bf J}_{\rm sc}}$
\end{enumerate}
%
%%%%%%%%%%%%%%%%%%%%%%%%%%%%%%%%%%%%%%%%%%%%%%%%%%%%%%%%%%%%%%%%%%%%%%%%%%%%%%%
%% FIGURE 1
%%%%%%%%%%%%%%%%%%%%%%%%%%%%%%%%%%%%%%%%%%%%%%%%%%%%%%%%%%%%%%%%%%%%%%%%%%%%%%%
\begin{figure}[!]
 \begin{center}\includegraphics[width=0.35\textwidth]{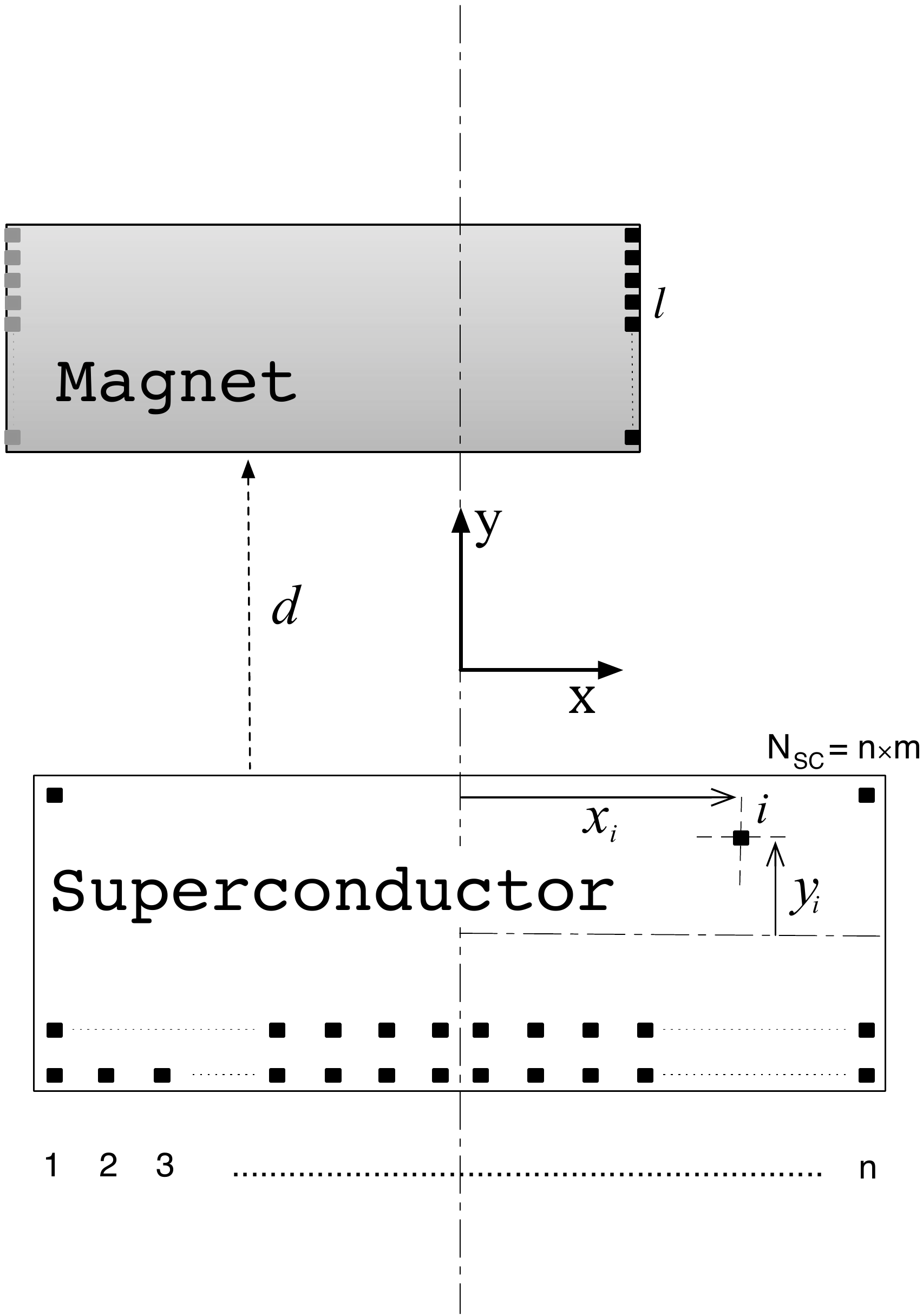}\end{center}
\caption{Finite element description of the electromechanical problem solved by the App. The superconductor is covered by a 2D mesh ($i=1,2,\dots N_{\rm sc}$), and the magnet by surface elements $l=1,2,\dots N_{\rm mag}$.}
\label{fig:1}       
\end{figure}
%
%%%%%%%%%%%%%%%%%%%%%%%%%%%%%%%%%%%%%%%%%%%%%%%%%%%%%%%%%%%%%%%%%%%%%%%%%%%%%%%
%
The material properties are dealt with as follows. The superconductor is considered a {\em so-called} hard  material, modelled through the {\em critical state} hypothesis\,\cite{gcs,bean}. In physical terms, this means that magnetic field variations will be countered by shielding currents, theoretically obtained by application of Faraday's law of electromagnetism, and the condition of being bounded by a certain maximum value. This is the so-called, critical current density ${J}_c$, the fundamental parameter of our problem. One can show that, wherever induced, shielding currents take the critical value , i.e. : $\|{J}({x,y,z})\|={J}_c$.

As for the magnetic material, an ideal PM behaviour will be assumed. This means that, irrespective of the interaction with the superconductor, a uniform invariant magnetization structure (source of ${\bf A}_{0}$) occurs.
\subsection{The physical statement: finite elements (FE)}
\label{subsec:2-1}
Next, we apply the above ideas to the specific configuration sketched in Fig.~\ref{fig:1}. A long permanent magnet is close to a long superconductor, both having a rectangular cross section and being parallel along the $z-$axis, as displayed. The magnet is assumed to be uniformly magnetized along the vertical direction. This may well serve as a model for a typical levitation experiment. More specifically, it fits the {\em levitation train} track geometry. It may also catch many facts about the forces between cylindrical magnets and superconductors within the coaxial configuration \cite{ref3}.

The evolution of the superconductor/permanent magnet structure is obtained in terms of current density functions ${\bf J}_{\rm sc}(\vec{r},t)$ and \textcolor{black}{${\bf J}_{\rm mag}(\vec{r})$}\textcolor{black}{. Respectively, they correspond} to the superconductor's macroscopic current density, and to the conventional {\em effective magnetization current density} for the case of the magnet\,\cite{M}. 
Owing to the necessity of using numerical techniques for solving the 2D problem stated above, we have introduced a finite element mesh as the support of the physical variables. As shown in Fig.~\ref{fig:1}, we describe the problem as a collection of elementary long wires, each characterized by a certain current density, i.e.:
\begin{equation}
j_{k}\equiv {J}_{z}(x_{k},y_{k})\, ,
\end{equation}
with $k=1,2,\dots N_G$ labelling each wire of the problem. For simplicity, $j_{k}\,'s$ will be considered as the components of a column vector, and named after Dirac's {\em ket} and {\em bra} notation. $N_G$ indicates the full number of current elements: $N_G=N_{\rm mag}+N_{\rm sc}$ with $N_{\rm mag}$ for the magnet, and $N_{\rm sc}$ for the superconductor.
\begin{eqnarray}
\ket{j} \equiv \left(
\begin{array}{c}
j_{1}\\
j_{2}\\
\cdot\\
\cdot\\
j_{N_G}
\end{array}
\right)\qquad , \qquad
\bra{j} \equiv (j_{1},j_{2},\dots,j_{N_G}) \, .
\end{eqnarray}

\textcolor{black}{Here, we recall that the concept of ``full set'' of current elements grouped in $\ket{j} $ is useful insofar as global calculations are considered. For instance, evaluation of the total magnetic field will be done by a linear operation (Biot-Savart's law) with $\ket{j} $. Nevertheless, as described below, a part of $\ket{j} $, viz. the magnetic component will be a constant array (rigid PM structure), whereas the superconducting part will be the set of unknowns to be solved for a given process.}

\subsection{The physical statement: variational principle}
\label{subsec:2-2}
{\paragraph{Description of the permanent magnet}
The behaviour of the permanent magnet may be introduced in a variety of forms. As said above, here, we recall that magnetic field effects are equivalently described either in terms of magnetic moments per unit volume (${\bf M}$) or by their effective magnetization currents. For the case of uniform magnetization considered here, they appear in the form of surface currents \cite{M}
\begin{equation}
{\bf J}_{\rm M}={\bf M}_{0}\times \, \hat{\bf n}=\pm\, {M}_{0}\,\hat{\bf z}
\end{equation}
locally, at the magnet's lateral sides (recall that $\hat{\bf n}$ stands for the unit vector normal to the surface). Then, in proper units the FE magnetic current density vector is given by
\begin{equation}
\ket{j_{\;\rm mag}}=\pm m_{0}\ket{\mathbb{1}}\, ,
\end{equation}
with $m_{0}=constant$ and $\ket{\mathbb{1}}$ standing for the column vector with $N_{\rm mag}$ ones. The associated vector potential at any point may be obtained by a linear relation\,\cite{ref3}
\begin{equation}
\ket{A_0}={\tt P} \ket{\mathbb{1}} \, .
\end{equation}
Here $A_0$ is a notation for the $z-$component of the vector potential (${\bf A}_{0}={A}_{0}\,\hat{\bf z}$) and $\tt P$ is a geometrical matrix connecting the source (magnet) and field points. Indeed, this involves the superposition of $N_{\rm mag}$ long wire expressions.
%
%%%%%%%%%%%%%%%%%%%%%%%%%%%%%%%%%%%%%%%%%%%%%%%%%%%%%%%%%%%%%%%%%%%%%%%%%%%%%%%
%% FIGURE 2
%%%%%%%%%%%%%%%%%%%%%%%%%%%%%%%%%%%%%%%%%%%%%%%%%%%%%%%%%%%%%%%%%%%%%%%%%%%%%%%
%
\begin{figure*}[t]
 \begin{center}
 \includegraphics[width=0.95\textwidth]{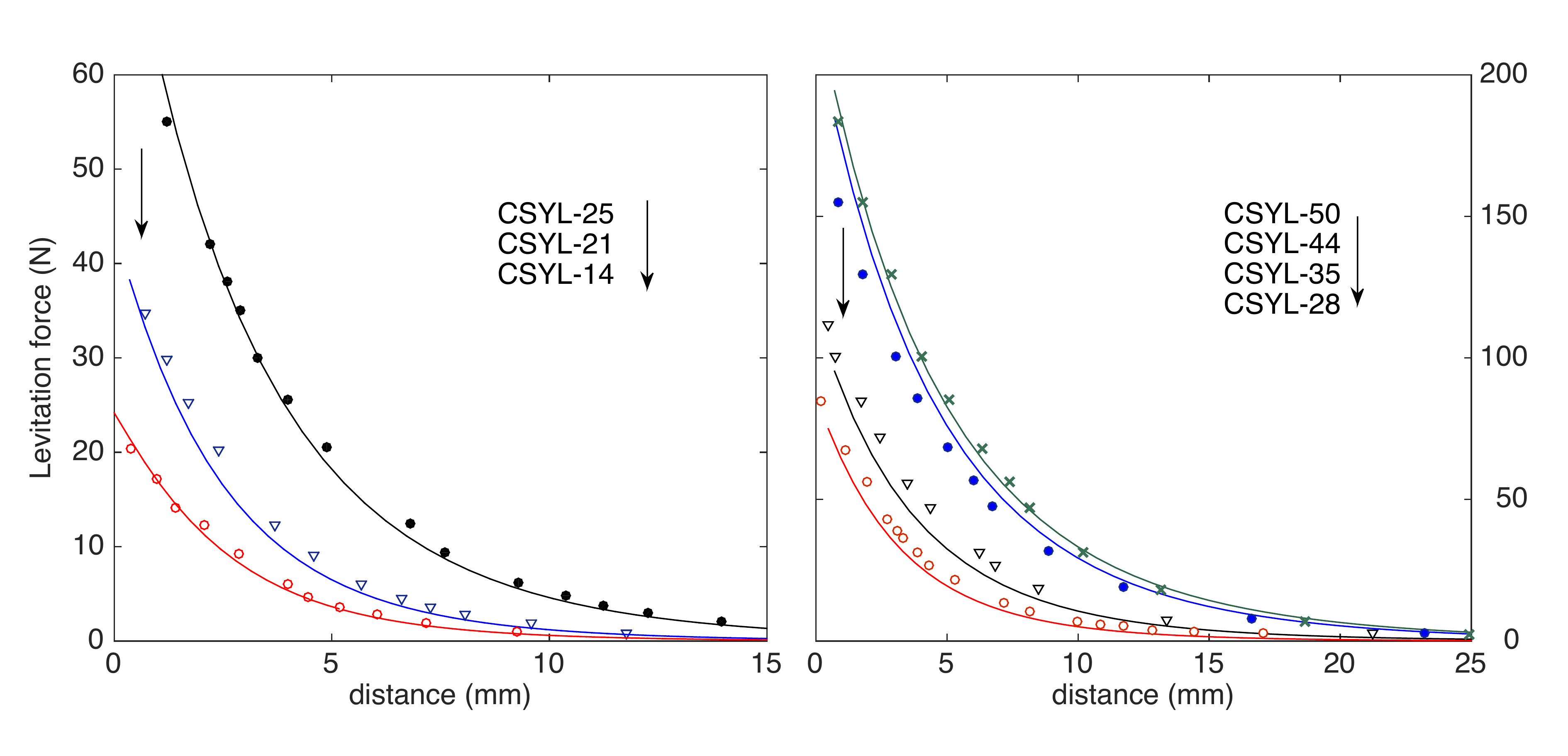}
 \end{center}
\caption{Comparison of model calculations (lines) vs. experimental data (symbols) for a variety of configurations, involving different magnets and superconducting samples. Naming convention reflects the diameter (mm) for the different superconducting samples (experiments taken from \cite{datacan}).}
\label{fig:2}     
\end{figure*}
%%%%%%%%%%%%%%%%%%%%%%%%%%%%%%%%%%%%%%%%%%%%%%%%%%%%%%%%%%%%%%%%%%%%%%%%%%%%%%%
%
{\paragraph{Superconductor's variational principle.}
In the FE formulation, the evolution of our simulation is defined by the following process

\begin{enumerate}
\item[(i)] The magnet is shifted, then $\ket{j_{\;\rm mag}}$ creates a new vector potential in space: ${A}_{0}^{n}\to {A}_{0}^{n+1}$
\item[(ii)] The superconducting currents have to be updated to the new situation $\ket{j_{\;\rm sc}^{(n)}}\to\ket{j_{\;\rm sc}^{(n+1)}}$. This is done by minimizing the following expression of the free energy\,\cite{ref3,gcs}

\begin{eqnarray}
\label{eqnFreeEmatrix}
{\tt F}[\ket{j_{\;\rm sc}^{(n+1)}}]=&&\frac{1}{2}\bra{j_{\;\rm sc}^{(n+1)}}{\tt M}\ket{j_{\;\rm sc}^{(n+1)}}
\nonumber
\\
&&-\bra{j_{\;\rm sc}^{(n)}}{\tt M}\ket{j_{\;\rm sc}^{(n+1)}}
\nonumber
\\
&&+\braket{{A_0^{(n+1)}-A_0^{(n)}}|{j_{\;\rm sc}^{(n+1)}}}
\nonumber
\\
\end{eqnarray}
under the restriction $|j_i| \leq j_c\quad\forall\; i=1,\dots N_{\rm sc}$.
\end{enumerate}

{In the above equation ${\tt M}$ represents the $N_{\rm sc}\times N_{\rm sc}$ mutual inductance matrix, between the superconducting elements, and $\bra{{A_0^{(n+1)}-A_0^{(n)}}}$ is the $1\times N_{\rm sc}$ row vector, formed by the increment of applied vector potential at the grid points of the superconductor.}

To be specific, a certain trajectory of the magnet is solved by starting with the condition $\ket{j_{\;\rm sc}^{(0)}}=\ket{{\mathbb 0}}$ for the superconductor\footnote[2]{This condition corresponds to the ``perfect conductor'' hypothesis, that is a good approximation for strong pinning materials in high magnetic fields} and successively ``updating'' the vector $\ket{j_{\;\rm sc}^{(n)}}$ for each step of displacement, by solving Eq.(\ref{eqnFreeEmatrix}).

Recall that (\ref{eqnFreeEmatrix}) is a ``quadratic problem'' subject to inequality constraints. In general, considering that the number of unknowns may be elevated (at least several hundreds of grid points for reasonable resolution) this requires a moderately high computational power. Just to give an idea, with an acceptable resolution, ``simulating a trajectory'' that entails displacements of a few centimetres in a typical levitation system consumes a bit more than 10 minutes in a standard desktop personal computer.
\subsection{Overview of the electromechanical problem}
\label{subsec:2-3}

The physical counterpart of Eq.(\ref{eqnFreeEmatrix}) is that, being exposed to variations of the ambient magnetic field, the superconductor reacts with a current density distribution (the finite element vector  $\ket{j_{\;\rm sc}}$) consisting of  the collection of values $j_i = \pm j_c , 0$ that better shields such variations. 
As it will be shown below, all the quantities of interest (forces, magnetic field lines, etc) may all be readily evaluated in terms of the applied magnetic field ${\bf H}_0$ (or its sources $\ket{j_{\;\rm mag}}$) and the magnetic induction created by the superconductor ${\bf B}_{\rm sc}$ (or its sources $\ket{j_{\;\rm sc}}$). 
Thus, starting with the Lorentz's force expression, each ``filament'' of magnetization current is subject to the force (per unit volume)
\begin{equation}
{\bf F}_{l}={\bf J}_{l}\times{\bf B}_{\rm sc}\, ,
\end{equation}
with ${\bf B}_{\rm sc}$ the superconductor's magnetic induction at the point $(x_{l},y_{l})$.

By using the notation introduced above, the total force per unit length of the magnet may be expressed
\begin{eqnarray}
\label{eq:forces}
f_x &&= -\braket{j_{\rm mag}|{b_y^{\rm sc}}}\equiv \bra{j_{\rm mag}}{\tt Q}_y\ket{{j}_{\rm sc}}
\nonumber\\
f_y &&=\quad \braket{j_{\rm mag}|{b_x^{\rm sc}}}\equiv \bra{j_{\rm mag}}{\tt Q}_x\ket{{j}_{\rm sc}}\, ,
\end{eqnarray}
with ${\tt Q}_{x,y}$ the geometrical matrices that couple the magnetic source and field points (superposition principle applied to the superconducting current elements here).
\subsection{Validation of the model}
\label{subsec:2-4}

Fig.~\ref{fig:2} shows the comparison of the simulation results obtained with our critical state finite element model and the experimental data corresponding to a set of measurements for superconducting bulks (YBaCuO) and permanent magnets (NdFeB) of different sizes \cite{datacan}. Experimental data were obtained by recording the force when moving the magnet towards the zero field cooled superconductor. Such a situation was recreated in our simulation by starting with the magnet at a big enough distance and the initial state $\ket{j_{\;\rm sc}}=\ket{{\mathbb 0}}$ in the superconductor.
We want to stress that a collection of noticeably different magnet and superconductor sizes are fairly reproduced in terms of a single \textcolor{black}{parameter}, i.e.: the critical current density. In all cases, we used $J_c = 3\cdot 10^8 A/cm^2$, \textcolor{black}{the value that better fits the experimental data and} that is reasonably within the expectations for these materials.

\textcolor{black}{As a further test of the model's predictive power, it must be mentioned that field cooled experiments and hysteresis of the levitation force were successfully reproduced in previous work\cite{ref3}.}
 %
%%%%%%%%%%%%%%%%%%%%%%%%%%%%%%%%%%%%%%%%%%%%%%%%%%%%%%%%%%%%%%%%%%%%%%%%%%%%%%%
%% FIGURE 3
%%%%%%%%%%%%%%%%%%%%%%%%%%%%%%%%%%%%%%%%%%%%%%%%%%%%%%%%%%%%%%%%%%%%%%%%%%%%%%%
%
\begin{figure}[!]
 \begin{center}
 \includegraphics[width=0.5\textwidth]{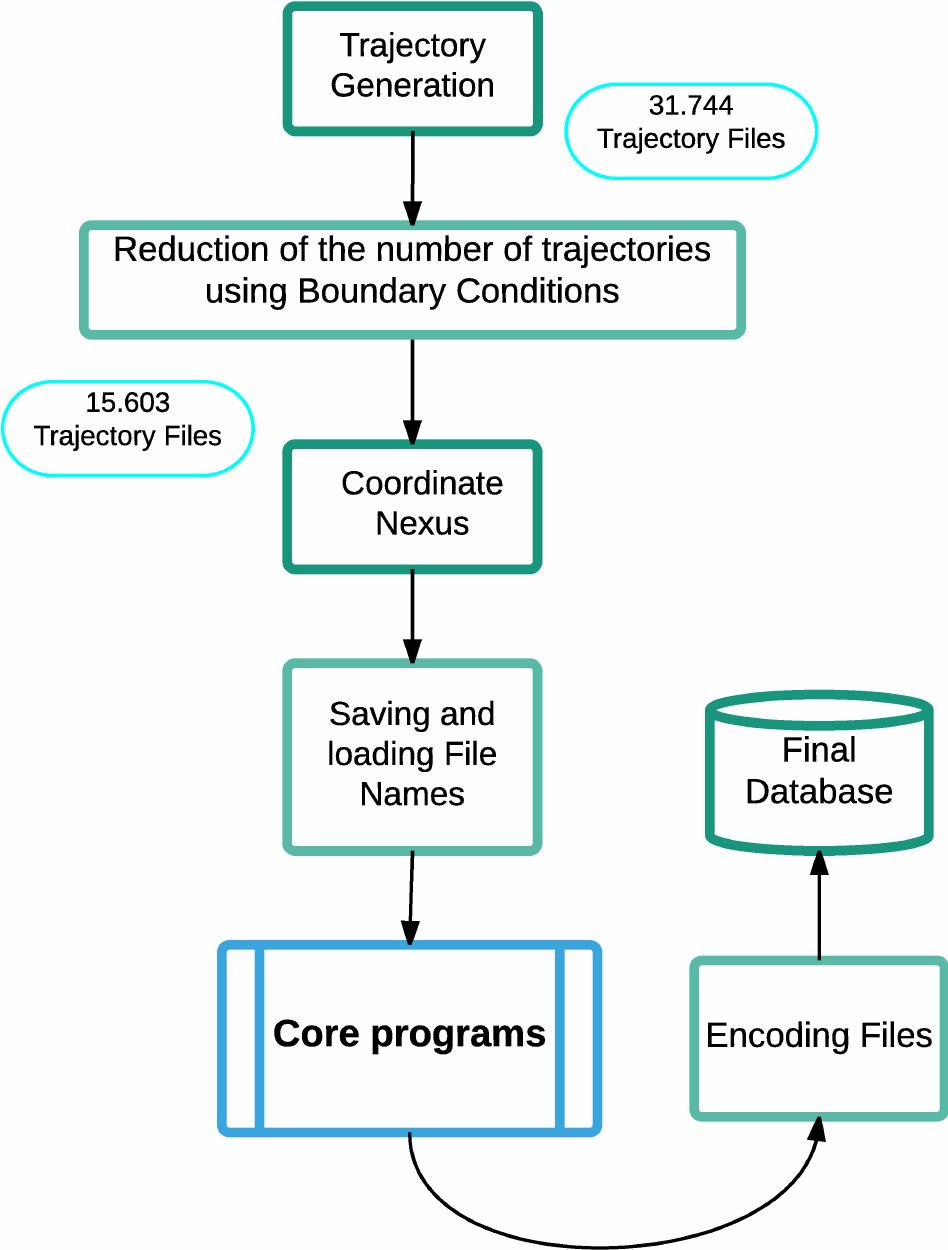}
 \end{center}
\caption{Sketch of the database setting-up.}
\label{fig:3}     
\end{figure}
%%%%%%%%%%%%%%%%%%%%%%%%%%%%%%%%%%%%%%%%%%%%%%%%%%%%%%%%%%%%%%%%%%%%%%%%%%%%%%%

\section{Notes on the implementation}
\label{sec:3}

Below, we describe our solution for performing a ``real-time'' simulation of the levitation experiment in the portable device. Details about software codes and solutions will be omitted here, though made available upon request for discussion with the interested reader. The whole set of computing resources is based on open-source modules.

As said before, computing the physical quantities that characterize the behaviour of the Maglev system, requires a combination of time consuming non-linear operations, and much faster linear post-processing evaluations. Among a number of possibilities, we have preferred the cooperation of intensive HPC (High Performance Computing) with the eventual operation of the portable device, that offers a good compromise. As sketched in Fig.\ref{fig:3}, it works under the plan:
\begin{enumerate}
\item[(i)] We define \textcolor{black}{a (large) finite number of trajectories which can be dealt with and that will constitute the set that may be investigated} by the end user.
\item[(ii)] The HPC performs the non-linear operations for the whole set (may take a long time).
\item[(iii)] Results are encoded and compressed to a data-base.
\item[(iv)] The database and post-processing codes are compiled to the App resources.
\item[(v)] Upon the user's request the App will read the data-base, perform post-processing calculations and visualize results in a virtual ``real-time'' evaluation.
\end{enumerate}

Let us briefly go through the main parts of this process.
%
%%%%%%%%%%%%%%%%%%%%%%%%%%%%%%%%%%%%%%%%%%%%%%%%%%%%%%%%%%%%%%%%%%%%%%%%%%%%%%%
%% FIGURE 4
%%%%%%%%%%%%%%%%%%%%%%%%%%%%%%%%%%%%%%%%%%%%%%%%%%%%%%%%%%%%%%%%%%%%%%%%%%%%%%%
%
\begin{figure}[t]
 \begin{center}
 \includegraphics[width=0.45\textwidth]{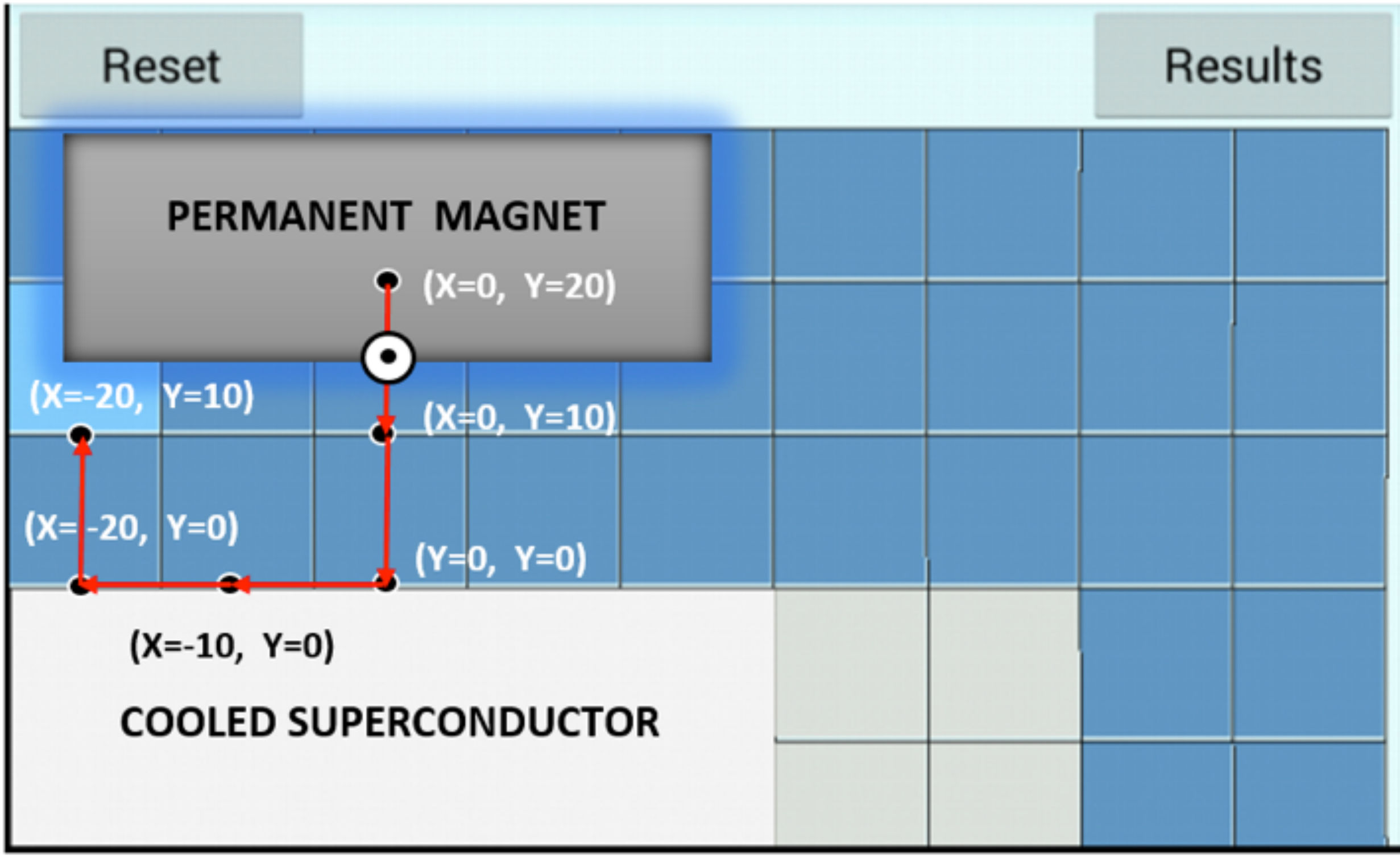}
 \end{center}
\caption{Definition of a trajectory for the magnet in the main screen of the App.}
\label{fig:4}      
\end{figure}
 %%%%%%%%%%%%%%%%%%%%%%%%%%%%%%%%%%%%%%%%%%%%%%%%%%%%%%%%%%%%%%%%%%%%%%%%%%%%%%%

\subsection{Simulation board, paths and trajectories}
\label{subsec:3-1}

As shown in Fig.~\ref{fig:4} the main screen of the App is conceived as a board of cells that may be successively occupied by the magnet. The superconductor remains rigid at the lower part. 

The end user will select a given trajectory, to be studied, just by tapping on the magnet and then sliding the finger along a set of neighbouring cells. 

Focusing on a typical geometry for experiments that are conducted in different instances as demonstrators and the characterization of materials for levitation machines, we have defined the following setup (as in Fig.~\ref{fig:4}), that fits the dimensions of commercially available materials. 
A permanent magnet of cross section $14\times 40\, mm$ is placed in the vicinity of a superconducting bar of cross section $14\times 50\, mm$. A board of 31 cells is defined and arranged in a {\em landscape} disposition for optimum visualization. 

A ``trajectory'' of the magnet is defined by a collection of coordinates for its centre. Each cell of the board corresponds to a $10\times 10\, mm$ square, and the centre of the magnet is allowed to jump in between any two consecutive cells, when defining a trajectory.

It is important to clarify that the discretisation described above relates to the definition of the ``trajectory'' across the board. Later, this will be rendered through the database that stores information of a much more complete ``path'' as calculated in a remote HPC server. Essentially, each step within the trajectory, defined by the sequence of positions on the board, unfolds in 10 substeps of real displacement (each of  $1\, mm$, correspondingly). This defines what we call the ``path''.
%
%%%%%%%%%%%%%%%%%%%%%%%%%%%%%%%%%%%%%%%%%%%%%%%%%%%%%%%%%%%%%%%%%%%%%%%%%%%%%%%
%% FIGURE 5
%%%%%%%%%%%%%%%%%%%%%%%%%%%%%%%%%%%%%%%%%%%%%%%%%%%%%%%%%%%%%%%%%%%%%%%%%%%%%%%
%
\begin{figure}[t]
 \begin{center}
\includegraphics[width=0.5\textwidth]{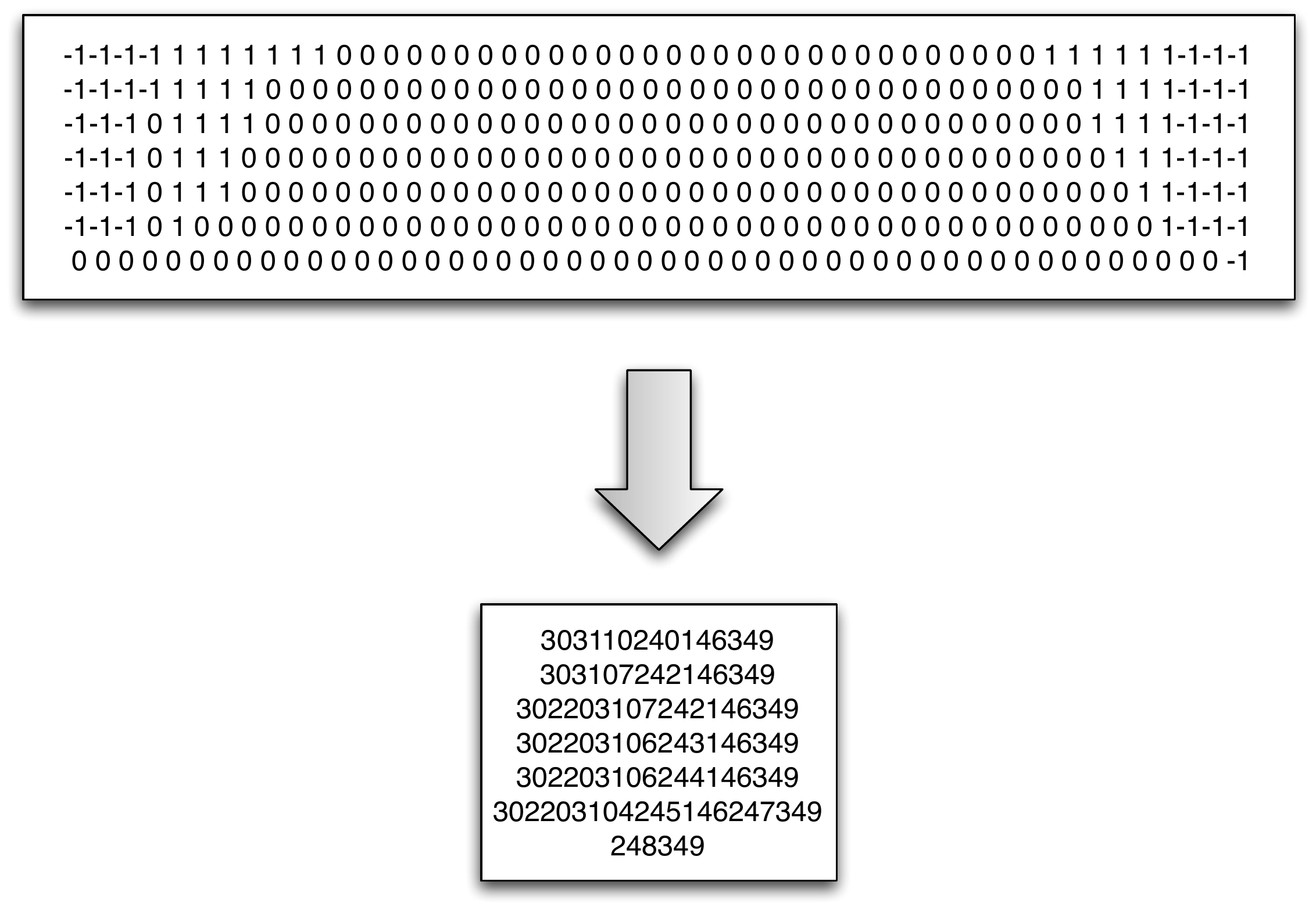}
 \end{center}  
\caption{Example of the numerical representation of the current density penetrating the superconductor. $\pm 1, 0$ indicate the current density $\pm J_{c},0$ at the given position. The lower pane shows a possible compression of this information.}
\label{fig:5}    
\end{figure}
%%%%%%%%%%%%%%%%%%%%%%%%%%%%%%%%%%%%%%%%%%%%%%%%%%%%%%%%%%%%%%%%%%%%%%%%%%%%%%%

\subsection{Encoded electromagnetic solution}
\label{subsec:3-2}

Simple combinatory and reduction by symmetry shows that our $31$ cell board gives way to $15603$ possible trajectories if a (reasonable) sequence of $5$ successive steps are allowed for the magnet starting at any position. Rather obviously, this amounts a lot of information and some care has to be taken. Starting from an initial size \textcolor{black}{of the output database file} of $\approx 4$\,GB, we could reduce it to  $\approx 40$\,MB. \textcolor{black}{This was done by removing redundant data in our initially sparse files.} Outstandingly, the main point of the compression was to consider the physical process itself.
As illustrated in Fig.~\ref{fig:5}, for each position of the magnet along a given path, what one must store is a collection of values for $\ket{j_{\;\rm sc}}$. However, as each component of this vector may only take one out of three values (say $\pm 1, 0$ in certain units) what one stores is a matrix as shown in the upper part of the figure. The full path is nothing but a pile of such matrices. But, most importantly, matrices may be further reduced in size by the simple trick of {\em codifying} the values $\pm 1, 0$ and just saving a counter for the number of each of these values along each row (lower pane of Fig.~\ref{fig:5}). \textcolor{black}{Needless to say, eventual interpretation of the so-compressed database requires to {\em decode} the information by the inverse of the compression algorithm, and will be done by the App software.}
%
%%%%%%%%%%%%%%%%%%%%%%%%%%%%%%%%%%%%%%%%%%%%%%%%%%%%%%%%%%%%%%%%%%%%%%%%%%%%%%%
%% FIGURE 6
%%%%%%%%%%%%%%%%%%%%%%%%%%%%%%%%%%%%%%%%%%%%%%%%%%%%%%%%%%%%%%%%%%%%%%%%%%%%%%%
%
\begin{figure}[t]
\hspace{-0.75cm}
\includegraphics[angle=0,width=0.575\textwidth]{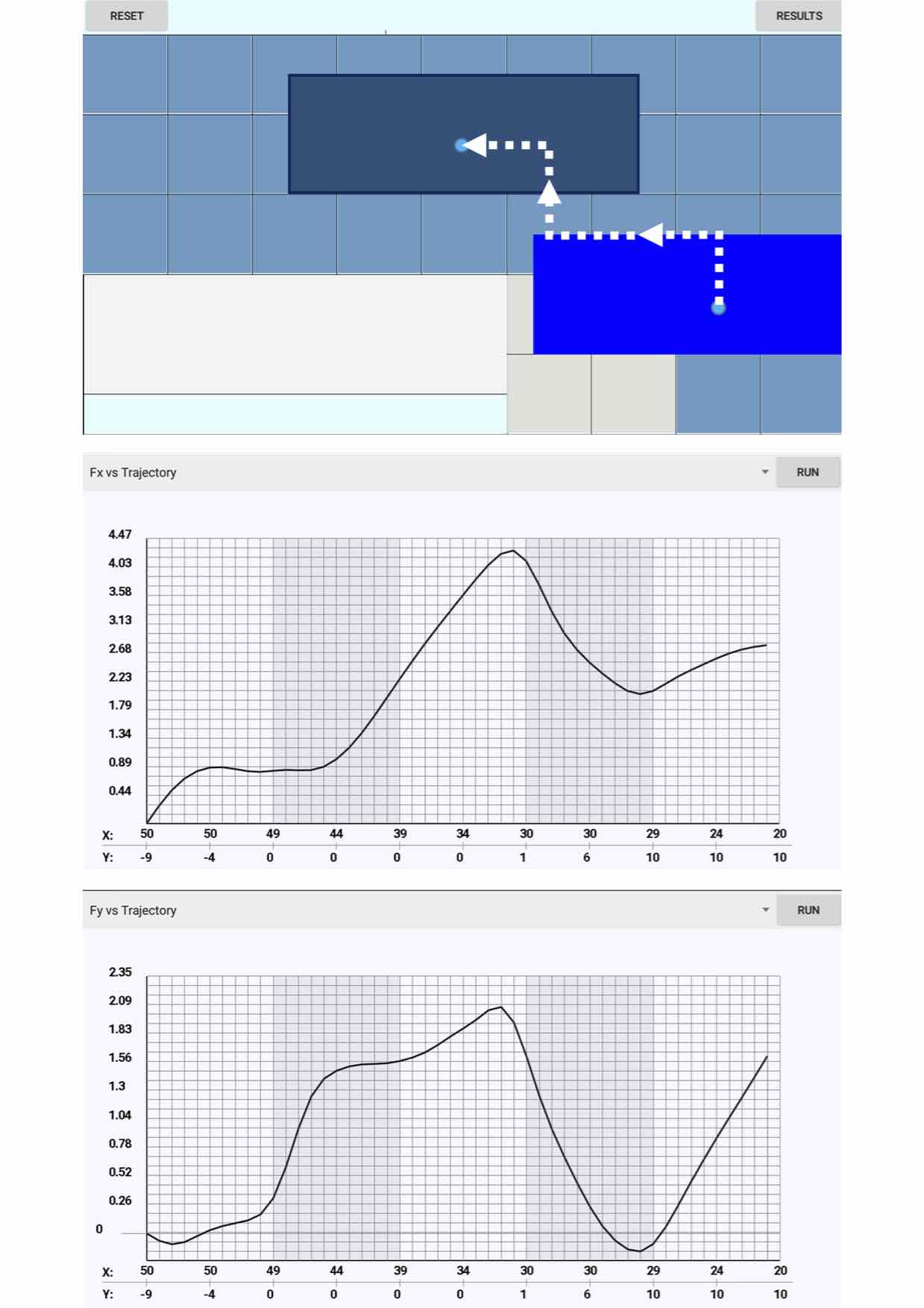}
\caption{Results of the simulation of the force vector components for a given trajectory of the magnet, as defined in the upper panel (snapshot of the screen of the device). \textcolor{black}{The middle panel shows the horizontal force component ($F_x$), and the lower panel displays the vertical force ($F_y$). The force units are $N$, and displacements given in $mm$.}}
\label{fig:6}      
\end{figure}
%%%%%%%%%%%%%%%%%%%%%%%%%%%%%%%%%%%%%%%%%%%%%%%%%%%%%%%%%%%%%%%%%%%%%%%%%%%%%%%%
%
\newpage
%%%%%%%%%%%%%%%%%%%%%%%%%%%%%%%%%%%%%%%%%%%%%%%%%%%%%%%%%%%%%%%%%%%%%%%%%%%%%%%
%% FIGURE 7
%%%%%%%%%%%%%%%%%%%%%%%%%%%%%%%%%%%%%%%%%%%%%%%%%%%%%%%%%%%%%%%%%%%%%%%%%%%%%%%
%
\begin{figure*}[tbh]
\begin{center}
\includegraphics[width=0.8\textwidth]{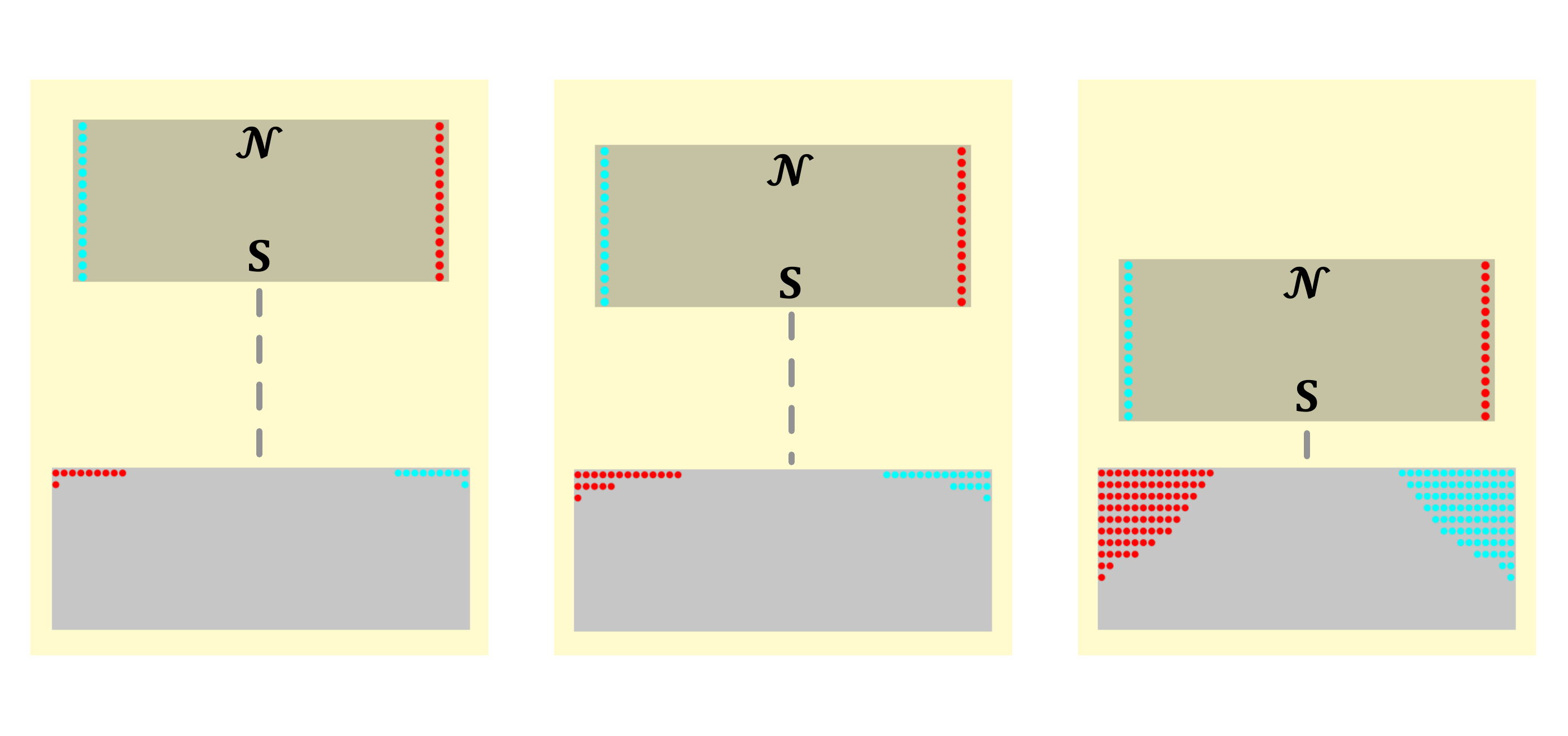}
\end{center}
\caption{Quick physical simulation: representation of the effective (magnet), and induced (superconductor) current densities for a ``centred'' trajectory in which the magnet approaches the superconductor. The {\em north} and {\em south} poles of the magnet are labelled together with the system of (given) effective surface currents.}
\label{fig:7}    
\end{figure*}
%%%%%%%%%%%%%%%%%%%%%%%%%%%%%%%%%%%%%%%%%%%%%%%%%%%%%%%%%%%%%%%%%%%%%%%%%%%%%%%
%

\subsection{Post-processing: evaluation of forces}
\label{subsec:3-3}

As explained above, the non-linear physical model for the levitation system is solved by means of an HPC facility, and the full information for a number of codified trajectories stored within a database. Eventually, the database has to be loaded in the end user's portable device that will perform post-processing steps involving linear operations on the data. 
\textcolor{black}{With current technologies, transferral of the data to the device takes several seconds with a direct cable connection, and usually much more through wireless communication. Thus, we suggest a single-operation for loading the software and data, and further stand-alone use of the App.}
In brief, eventual data manipulation for the calculation of forces, magnetic fields, etc is based on matrix multiplication operations, carried out by the device ({\sc Android} system in our case)

The calculation of the force components arising for a given trajectory of the magnet (see Sec.\ref{subsec:2-3}) is illustrated in Fig.~\ref{fig:6}. 
The relatively simple operations involved in Eq.(\ref{eq:forces}), are efficiently performed, \textcolor{black}{even so the matrix elements ${\tt Q}_{x,y}(k,l)$ are evaluated in real-time during the calculations}

 Upon selection of the trajectory by the user, the device searches the database, identifies the collection of induced superconducting currents $\ket{{j}_{\rm sc}}^{\rm (1)}\!,\ket{{j}_{\rm sc}}^{\rm (2)}\!,\dots$, applies Eq.(\ref{eq:forces}) and displays results as shown in Fig.~\ref{fig:6}. Notice that each of the 5 steps in the selection corresponds to one of the highlighted vertical bands in the plot of forces. The lower part of the figure displays the values of the $(x,y)$ coordinates of the magnet's center, so that one can immediately match with the screen showing the simulation board above. Recall that trajectories are defined relative to the position of the magnet centred on top of the superconductor. 

%Details about the interpretation of this information will be given in Sec.\ref{sec:4}.
%
 \textcolor{black}{As a final comment}, concerning Fig.~\ref{fig:6}, we notice that the displayed values are used to scale the axes automatically, establishing the range on the screen. Nevertheless, physical units for the force and displacement (newtons and millimetres) are used, corresponding to real values in the simulated system. In our case, assuming that both materials are in the range of best performance, a NdFeB permanent magnet with \textcolor{black}{remnant magnetization $\mu_{0}M_{0}=1.17\, T$}, and a melt textured superconductor with critical current density of $0.1\, GA/m^2$ were considered in the simulation.

\subsection{Visualization of induced current densities and magnetic field lines}
\label{subsec:3-4}

In addition to the quantitative analysis of the levitation ($F_y$) and guidance ($F_x$) forces\footnote[6]{\textcolor{black}{The concept of lateral ``guidance'' in a MagLev system is as follows. Stabilizing guidance action is related to the appearance of induced currents in the superconductor that will give place to restoring forces, i.e.: if the magnet is moved to the right (say through a displacement $x\to x+\delta x$) the arising force points to the left (say $F_{x}=-k\,\delta x$ with $k>0$)}}, the simulator incorporates the option of providing visual information. The user will obtain a trustworthy representation of the current profiles induced within the superconductor, as well as a picture of the magnetic field lines over the region of interest. As it has been traditionally exploited in Electromagnetics, this graphical information may be reassuring on the underlying calculations, and also serve as a helpful guide for comprehending the physical processes.

Plotting a representation of the induced current is a simple operation. For example, in our case, a red point is used to indicate positive current flow (${J}=+{J}_c$), whereas a blue one corresponds to negative flow (${J}=-{J}_c$). A blank is used for the areas free of current circulation (${J}=0$). 

When magnetic field lines are to be displayed, a somehow involved computation process is necessary. The portable device must perform calculations (as will be described below) and this may slow down the output of results noticeably. For this reason, it seemed convenient to  endow the App with the choice between a ``quick'' and a ``full'' simulator so that the end user can compromise between the amount of information needed and the time required to get it on a certain portable device.

\paragraph{Quick simulator}
With this option selected, the simulator merely shows a map of the current density penetration in the superconductor as the magnet moves. Effective surface currents in the permanent magnet have a frozen profile, whereas superconducting current densities display a certain dynamics as the magnet moves. For instance, starting from a distant position and lowering the magnet, one can observe a basically opposite magnet (something like a mirror image) induced in the superconductor. This has been checked in Fig.~\ref{fig:7}. The superconductor becomes a kind of effective magnet that opposes to the North/South pole structure of the real magnet, and thus gives way to a repulsion force between both.

More complex structures as those originated when the magnet is shifted laterally or oscillated may be observed and will be discussed in the next section.

\paragraph{Full magnetic field structure}
As introduced before, having obtained the \textcolor{black}{superconducting} current density distribution for each time step at a given trajectory of the permanent magnet, the full magnetic field structure follows by a linear analysis. This operation is performed by the portable device itself. The process is as follows.

The vector potential is evaluated as the finite element vector
\begin{equation}
\ket{A_z}={\tt M}\ket{j} \, .
\end{equation}
Here, ${\tt M}$ is the mutual inductance matrix coupling the full grid and the field sources, i.e.: $\ket{j} $ the direct sum of permanent magnet and superconducting current density finite element vectors. \textcolor{black}{As said in Sec.\ref{subsec:2-1} this is the $(N_{\rm mag}+N_{\rm sc})\times 1$ column vector that merges the position dependent constant components of effective magnetization currents and the updated distribution in the superconductor for each position of the magnet}:
\begin{equation}
\ket{j}=\ket{j_{\rm mag}}\oplus\ket{{j}_{\rm sc}} \, .
\end{equation}

Once $A_{z}(x,y)$ has been evaluated in the grid for a given time step, one may use the 2D property: ``the force lines of ${\bf B}$ coincide with the isolines of $A_z$'', which immediately follows from the relation

\begin{eqnarray}
\left.
\begin{array}{ll}
\displaystyle{B_x = \frac{\partial A_z}{\partial y}}\\
\\
\displaystyle{B_y = -\frac{\partial A_z}{\partial x}}
\end{array}
\right\}
\Rightarrow dA_z = -B_y dx + B_x dy\, ,
\label{eq:2dAB}
\end{eqnarray}
equivalent to
\begin{equation}
{\bf grad}{A_z}\cdot {\bf B} = 0\, ,
\end{equation}
meaning that the isolines of $A_z$ are parallel to ${\bf B}$.

\textcolor{black}{Thus, the visualisation of ${\bf B}$-lines relies on a contour plotting task. Dedicated computer software offers solutions to this problem, which is by no means trivial if one starts from the scratch. In our case, starting from the general concepts in Ref.\cite{isolines}, a ``low level'' solution was implemented in the {\sc Java} language that may be integrated in the {\sc Android} activity. For the readers' sake, the fundamental ideas of our solution are explained in \ref{sec:appendix}.}

%
%%%%%%%%%%%%%%%%%%%%%%%%%%%%%%%%%%%%%%%%%%%%%%%%%%%%%%%%%%%%%%%%%%%%%%%%%%%%%%%
%% FIGURE 8
%%%%%%%%%%%%%%%%%%%%%%%%%%%%%%%%%%%%%%%%%%%%%%%%%%%%%%%%%%%%%%%%%%%%%%%%%%%%%%%
%
\begin{figure}[!]
\begin{center}
\includegraphics[width=0.45\textwidth]{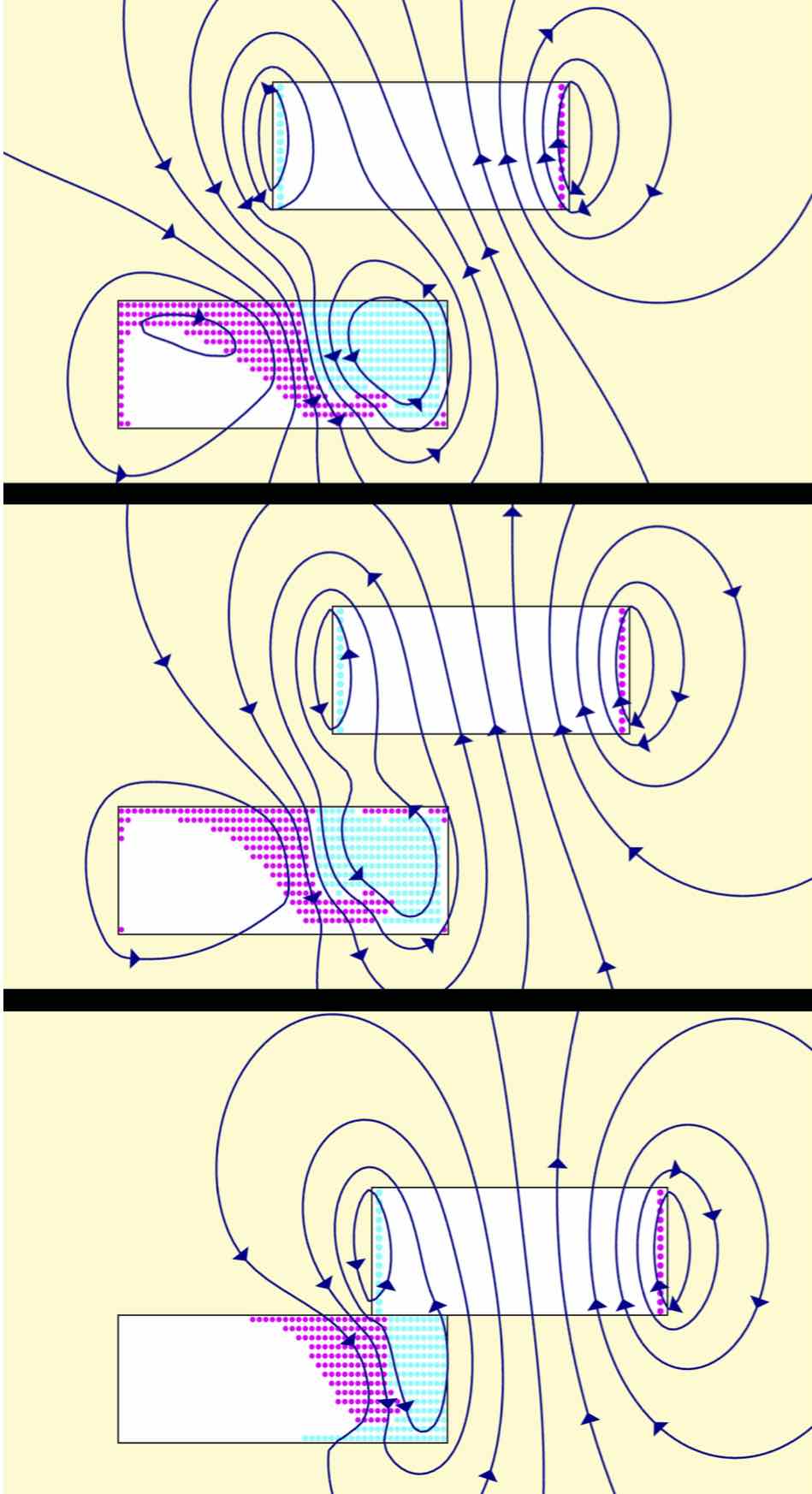}
\end{center}
\caption{Simulation of magnetic field lines in the vicinity of a permanent magnet/superconducting system for a \textcolor{black}{{\em zigzag} upwards trajectory} (snapshots of the trajectory defined in Fig.\ref{fig:6}).}
\label{fig:8}    
\end{figure}
%
%%%%%%%%%%%%%%%%%%%%%%%%%%%%%%%%%%%%%%%%%%%%%%%%%%%%%%%%%%%%%%%%%%%%%%%%%%%%%%%
%
\textcolor{black}{Here, just as an illustration of the above process, Fig.~\ref{fig:8} shows the induced electric currents and the magnetic field structure, corresponding to an upwards zigzag movement. To be specific, corresponding to the {\em trajectory} defined in Fig.\ref{fig:6} we show three snapshots of the refined {\em path} from one discrete cell to the other, corresponding to the finger positions $2.5\to 3.5\to 5$}

\newpage

\section{Concluding remarks}
\label{sec:4}

Up to date, numerical simulation techniques that take advantage of the high visualization potential and non-negligible computation capacity of portable devices have been barely touched. Not to mention is the intuitive, user-friendly environment typically developed for such machines. Here, a possible strategy for implementing the idea in a highly demanding sector has been investigated. A simulator for magnetic force levitation appliances have been developed. Specifically, we targeted on a stand-alone simulator that may be used as a simple modelling software by technologists, as well as a practical toolkit for material selection, experiment interpretation or even as a help tool in scientific discussion. \textcolor{black}{Training, dissemination and public engagement actions are certainly a focus of our software.}

Being \textcolor{black}{interested in} obtaining realistic results, as a feasible strategy for developing the simulator (ultimately running on portable devices) we have chosen the combination of High Performance Computing on a Linux cluster, and eventual simplified post-processing calculations on the {\sc Android} based system. The concept of database (transferred from the cluster to the device in a singular connection) has been essential. This allows to obtain the physical information, i.e.: induced electromagnetic fields and arising mechanical forces, in real time.

%Basically, the physical background of the simulator is the critical state model for the superconducting component. Comparison to experimental data has been performed and validates the realistic output of the software, for the physical processes considered. They comprehend a high number of trajectories for the magnet wagging around the superconductor, which are depictive of many relevant applications, that typically imply distances between the elements in the sub-centimetre range.
%
%In its present form the simulator evaluates the behaviour of a system composed by a magnetic bar with cross section $14\times 40\, mm$ in the vicinity of a superconducting bar of section $14\times 50\, mm$. The modelled magnetic material has the properties of NdFeB, with a remanence field $\mu_{0}H = 1.17\, T$, and the superconductor is assumed to be a high quality melt textured YBaCuO bar with the critical current density value $J_c = 0.1\, GA/m^2$. These are just representative values, but may be changed if necessary. In fact, by ``recreating'' the database, one may simulate the behaviour of any system of interest. Different sizes and materials could be taken into account, together with specific SC/PM assemblies with different geometry. In this sense, considering that our App was built by combination of open source modules, one may customize it for any given purpose. 

A number of improvements of the tool have been discussed. Among them, we want to mention that, in its present form, the simulator investigates the ``quasi-static'' interaction between the superconductor and a permanent magnet that describes a well defined trajectory imposed by some external action. In physical terms, we are \textcolor{black}{\em evaluating} the magnetic forces acting on the magnet along some virtual displacement. Such displacement is not necessarily equal to the one followed by the system after a perturbation of equilibrium. Strictly speaking, the simulation evaluates the response of the system to a definite action positioning the system. However, the real dynamics followed by the magnet in a maglev application would occur along a specific trajectory obtained by applying the least action principle with electromagnetic and gravitational interactions included. In addition, rotational torques (also ``absorbed'' by our external action here) should be included in the description of the system. 

\textcolor{black}{Another aspect that should be mentioned is that, being based on the {\em critical state model}, our simulations do not allow to consider time dependent relaxation effects related to the finite slope of the superconductor's current-voltage characteristic. Nevertheless, this is not really a limitation of our numerical tool. One could just generate a dedicated database by replacing the restriction $| j_i| \leq j_c$ in Eq.(\ref{eqnFreeEmatrix}) with a {\em penalty} term accounting for the related losses\cite{relax} and relaxation would be accounted.}

Continuing work along the above lines is intended.

\ack

Dr. F. L\'opez--Tejeira is gratefully acknowledged for many useful comments and suggestions along the development of this work.

Funding of this research by Spanish MINECO and FEDER programme (Project ENE2014-52105-R-9) and by Gobierno de Arag\'on (Research group T-12) is  gratefully acknowledged. J.  Pe\~{n}a-Roche acknowledges a Research Grant by I.C.M.A. (PI$^2$ programme 2015).

%\newpage

\begin{appendices}
\numberwithin{figure}{section}
\numberwithin{equation}{section}
\renewcommand{\theequation}{\Alph{section}.\arabic{equation}}
\renewcommand\thesection{Appendix \Alph{section}}
\renewcommand\thefigure{\Alph{section}\arabic{figure}}
\section{\textcolor {black}{Algorithm for generating the lines of force of the magnetic field}}
\label{sec:appendix}

The algorithm to plot the contours of $A_z$ (and, thus, to generate the field lines of ${\bf B}$) is based on Fig.~\ref{fig:A1}, and was implemented in the {\sc Java} language. It relies on the combination of a procedure that seeks the points at which the function $A_z(x,y)$ takes the desired values and the use of a {\em dual mesh} that allows to establish a classification of ordered pairs for plotting the contours\,\cite{isolines}. Basically, one starts by defining a desired family of $n_c$ level contours   
\begin{equation}
\{ L_k\, | \, A_z(x_k,y_k)=L_k\, , k = 1,2,\dots n_c\}
\end{equation}
and then identifies the sets of points  \textcolor{black}{$\{(x_k,y_k)\}$} in which the condition \textcolor{black}{$A_z=L_k$} is fulfilled. This is done by double loops along the grid points $(i,j)$, sweeping both vertically and horizontally. Most probably, just a small (or null) number of grid points will precisely correspond to the locus of a given contour, and thus, interpolation will be necessary. For instance, the points labelled $k_N , k_M$ in Fig.~\ref{fig:A1} are obtained by linear interpolation in between the grid points that give place to change of sign in the quantity $A_z-L_k$:
\begin{eqnarray}
&&A_{z}(i-1,j)-L_k <0
\nonumber\\
&&A_{z}(i,j)-L_k >0
\nonumber\\
&&A_{z}(i,j+1)-L_k <0\, .
\end{eqnarray}
Thus, each contour $L_k$ is defined by a set of indices
\begin{equation}
{\tt\bf k}\equiv \{ k_1 , k_2 ,\dots   k_n , k_m,\dots \}\, ,
\end{equation}
corresponding to a set of points, that are stored and later joined by segments.
The important issue about how to sort the points of the set ${\tt\bf k}$ so as to later join them properly is solved with help of the dual mesh. Two points of the set ${\tt\bf k}$ will be joined by a line if they share neighbouring squares. In our example, $k_n$ and $k_m$ have the common neighbour labelled ``10'' and so that, they are joined.
%
%%%%%%%%%%%%%%%%%%%%%%%%%%%%%%%%%%%%%%%%%%%%%%%%%%%%%%%%%%%%%%%%%%%%%%%%%%%%%%%
%% FIGURE A1
%%%%%%%%%%%%%%%%%%%%%%%%%%%%%%%%%%%%%%%%%%%%%%%%%%%%%%%%%%%%%%%%%%%%%%%%%%%%%%%
%
\begin{figure}[!]
\begin{center}
\includegraphics[width=0.35\textwidth]{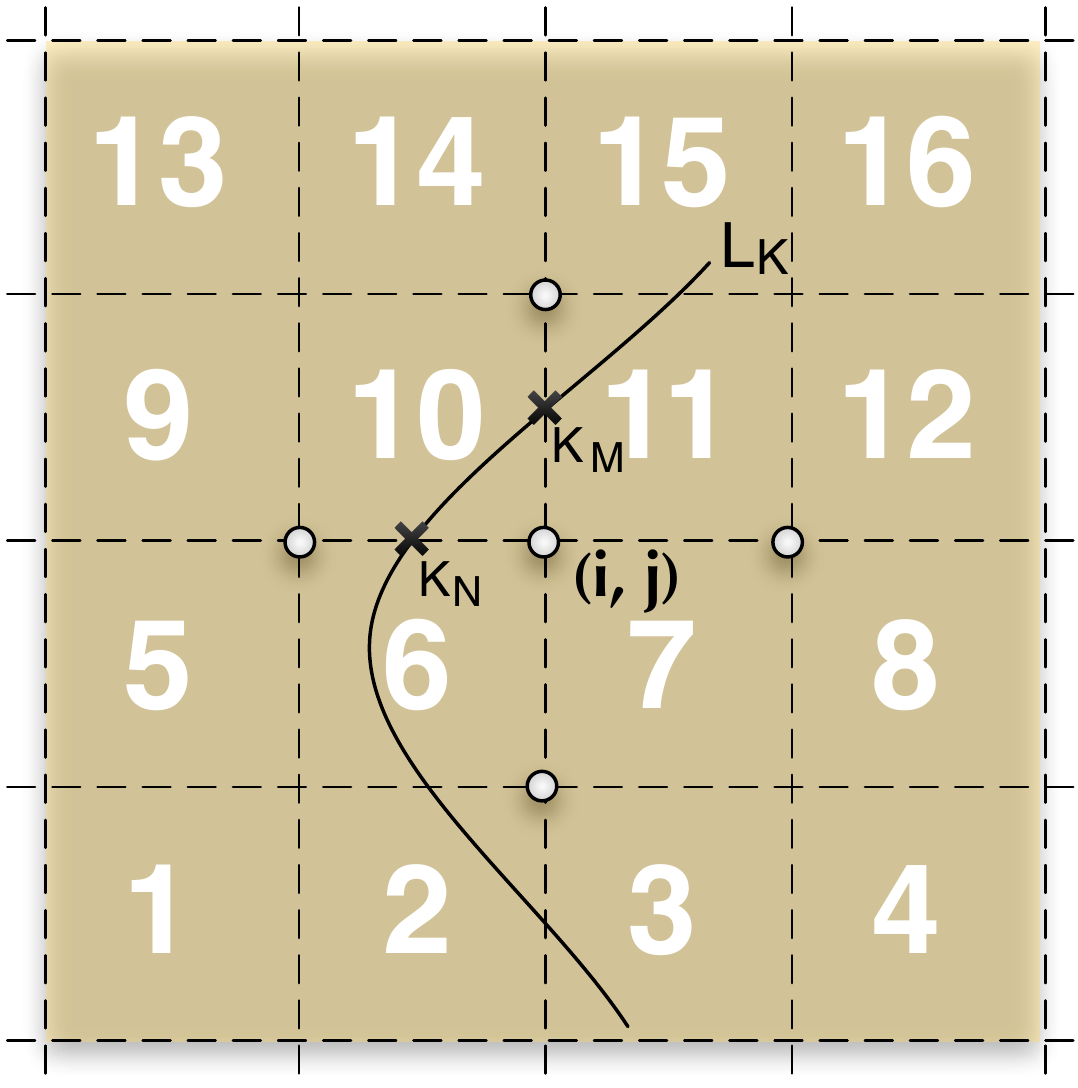}
\end{center}
\caption{Example of evaluation of a magnetic field line. $L_K$ defines a given contour line. $K_N$ and $K_M$ correspond to exact locations where the vector potential interpolates to the value $L_K$. $(i,j)$ is the point of the primal mesh. Numbered squares define a dual mesh.}
\label{fig:A1}    
\end{figure}
%
%%%%%%%%%%%%%%%%%%%%%%%%%%%%%%%%%%%%%%%%%%%%%%%%%%%%%%%%%%%%%%%%%%%%%%%%%%%%%%%
%
Following the above procedure, one can produce a plot on the screen that shows the lines of ${\bf B}$, but a further piece of information is still desirable. As distinction between ``positive'' and ``negative'' magnetic poles is important, the field lines must be oriented, for instance by means of arrows that indicate the sense of the field vector. Again, this property may be obtained from the landscape $A_z (x,y)$, by recalling the relation between $A_z$ and ${\bf B}$. In fact, using Eq.(\ref{eq:2dAB}) one has the particular relations
\begin{eqnarray}
&&d{\Bell}=(dx,0)\quad\Rightarrow\quad dA_z = -B_y dx
\nonumber\\
&&d{\Bell}=(0,dy)\quad\Rightarrow\quad dA_z = B_x dy\, .
\end{eqnarray}
{The orientation of the magnetic field} lines is obtained by comparison of the values of $A_z$ around a given point. More specifically, 
\begin{eqnarray}
{\rm sign}(B_x)&=&{\rm sign}(dA_z)\quad\quad\;{\tt for}\quad dy>0\; ,\; dx =0 
\nonumber\\
{\rm sign}(B_y)&=&-{\rm sign}(d{A}_z)\quad{\tt for}\quad dx>0\; ,\; dy =0 \, 
\end{eqnarray}
and this information allows to draw an oriented arrow for any given point of the contour line, just by comparing ${A}_z$ at the neighbouring points.

A simple algorithm for plotting a filled triangle at a chosen family of points has been implemented to perform this task.

\end{appendices}

\end{document}